\documentclass[aps,prl,twocolumn,groupedaddress]{revtex4-1}
\usepackage[pdftex,bookmarks,colorlinks]{hyperref}
\pdfoutput=1

\usepackage[pdftex]{graphicx}
\usepackage{natbib}

\begin{document}


\title{Order-to-disorder transition in ring-shaped colloidal stains}



\author{\'Alvaro G. Mar\'in}
\email[]{a.g.marin@utwente.nl}
\author{Hanneke Gelderblom}
\email[]{h.gelderblom@utwente.nl}
\author{Detlef Lohse}
\author{Jacco H. Snoeijer}


\affiliation{Physics of Fluids Group, Faculty of Science and Technology\\
Mesa+ Institute, University of Twente,\\
7500AE Enschede, The Netherlands.}



\date{\today}

\begin{abstract}
A colloidal dispersion droplet evaporating from a surface, such as a drying coffee drop, leaves a distinct ring-shaped stain. Although this mechanism is frequently used for particle self-assembly, the conditions for crystallization have remained unclear. Our experiments with monodisperse colloidal particles reveal a structural transition in the stain, from ordered crystals to disordered packings. We show that this sharp transition originates from a temporal singularity of the flow velocity inside the evaporating droplet at the end of its life. When the deposition speed is low, particles have time to arrange by Brownian motion, while at the end, high-speed particles are jammed into a disordered phase.
\end{abstract}

\pacs{}

\maketitle



%


Ordered arrays of colloidal particles present important characteristics for fields as photonics \cite{colloidalphotonics} and biotechnology \cite{evaporativeDNA}. A relatively simple approach to deposit particles onto a substrate is by evaporation of a colloidal dispersion droplet. In droplets with pinned contact lines, evaporation gives rise to the so-called coffee-stain effect \cite{Deegan:1997, Deegan:2000, Hu:2005, Popov:2005, bodiguel:2010, mugele}: a ring-like stain remains on the substrate once the liquid has evaporated. The mechanism behind this ring-stain formation is explained in the pioneering work of Deegan et al. \cite{Deegan:1997}: In an evaporating drop with an immobile contact line, a capillary flow is generated to replenish the liquid that has evaporated from the edges. This flow drags particles towards the contact line, forming the ring-shaped stain in particle suspensions such as coffee.

This deposition mechanism has been successfully used to generate colloidal crystals in three dimensions \cite{Velikov:2002}. Furthermore, techniques have been developed to control the patterns by permitting evaporation through specially designed masks \cite{Harris:2007} or by controlling the wettability \cite{KStebe2004assembly, mugele}. Also large-scale colloidal crystals have been observed for particle sizes below 100 nm \cite{Bigioni_Witten}, but some displayed instabilities that gave rise to concentric rings \cite{Pauchard}, stripes, chevrons \cite{Limat:1995}, and cracks \cite{Dufresne:2003}. At present, there is no way of predicting beforehand the structure of the deposit remaining after evaporation, due to a lack of understanding of the basic mechanisms. However, the processes driving colloidal crystallization are crucial for controlling the mechanical and the optical properties of the synthesized materials.

\begin{figure}
\includegraphics[width=3.4 in]{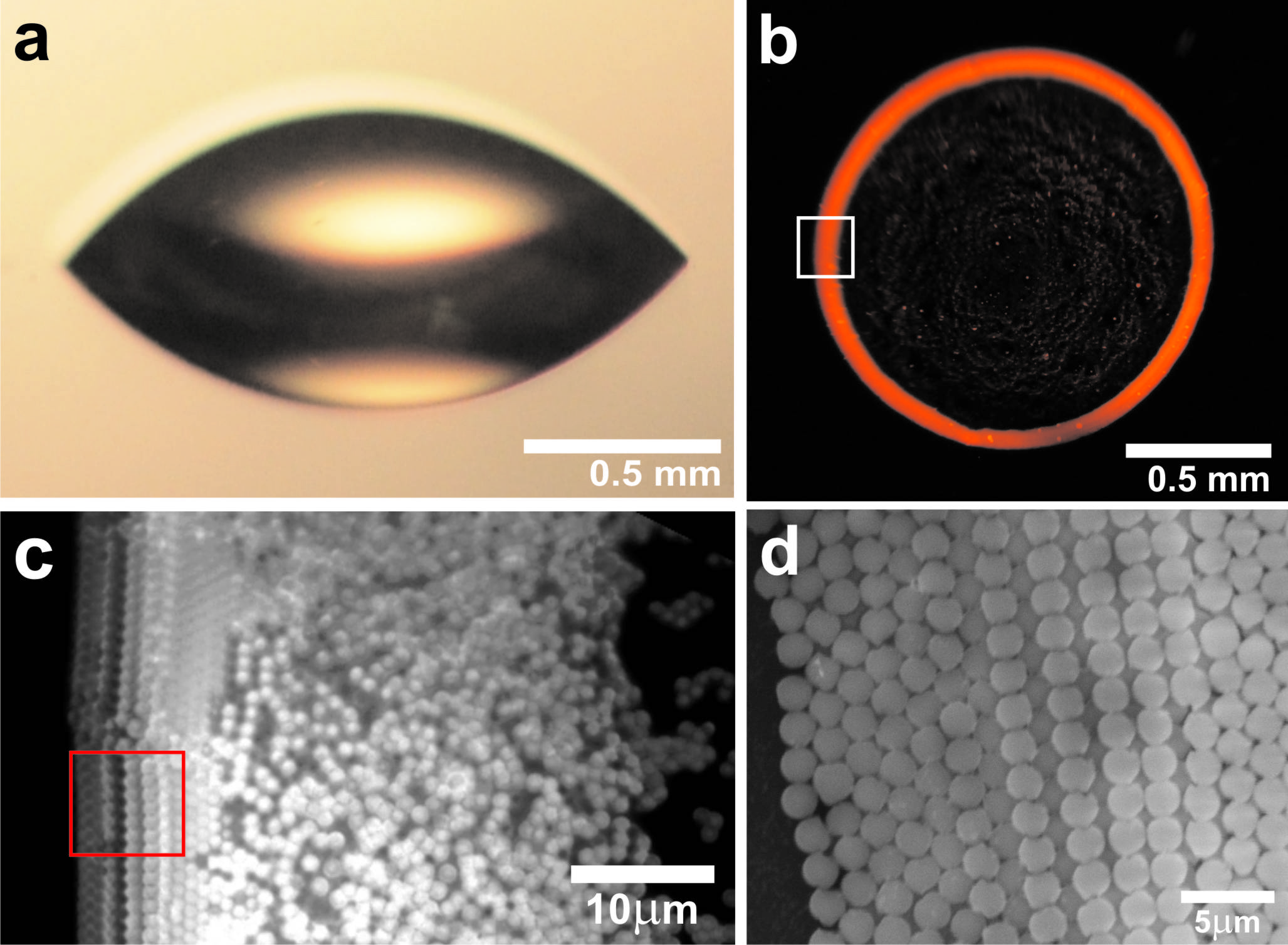}
	
\caption{Order-to-disorder transition in the particle stain left by an evaporating drop. (a) A 3-$\mu$L sessile water droplet evaporates from a glass substrate. (b) Ring-shaped stain of red particles (the coffee stain) left on the substrate after evaporation. (c) A close-up by an optical microscope of the bottom layer of the stain, taken from the white square in (b), shows that the outermost lines of the stain (left) have an ordered, crystalline structure. Towards the center of the drop (right), a transition to a disordered particle arrangement is observed. (d) A  top view of the ring stain, taken from the red square in (c) with the Scanning Electron Microscope (SEM), shows that the first lines of particles (left) are arranged in a hexagonal array, while the following lines (brighter in the image) are arranged as square, followed by again hexagonal array.}
\label{fig1}
\end{figure}

In this Letter, we reveal the conditions leading to colloidal crystallization in a paradigmatic system: a 3-$\mu$L sessile water droplet containing colloidal particles evaporating from a smooth glass slide (Fig. \ref{fig1}a). The colloids consist of almost monodisperse red-fluorescent polystyrene particles with particle diameters in the range of 0.5 to 2 $\mu$m. The droplet is gently deposited onto a glass, and is left to evaporate under controlled temperature and humidity. The contact line remains pinned during the entire experiment.
Our aim is to analyze structure of the particle arrangement in the remaining coffee stain. It turns out that the dynamics of particle deposition is crucial for the understanding of the final structure.
Therefore, the entire evaporative process is filmed simultaneously from the side and from below \cite{sup}. This allows to follow the evolution of the drop geometry, the formation of the deposit, and the dynamics of particle migration. The particles are filmed from below with a CMOS camera (PCO imaging) connected to an inverted microscope. An integrated standard halogen light source is used to illuminate the particles; the emitted light is filtered before it reaches the camera. The camera is operated at a typical frame rate of 1 frame per second, which is sufficient to capture the movement of the particles. To determine the velocity field in the measurement area from the captured images, a custom-made MATLAB micro-Particle Image Velocimetry ($\mu$PIV) code was used. The side-view images are used to calculate contact angle, volume, radius, and height of the drop at every instant with a custom-made MATLAB code. Side-view images are taken with a long distance microscope and a CCD camera (Lumenera Corp.), synchronized with the visualization of the particles by a common trigger.

\begin{figure}
	\includegraphics[width=3.4 in]{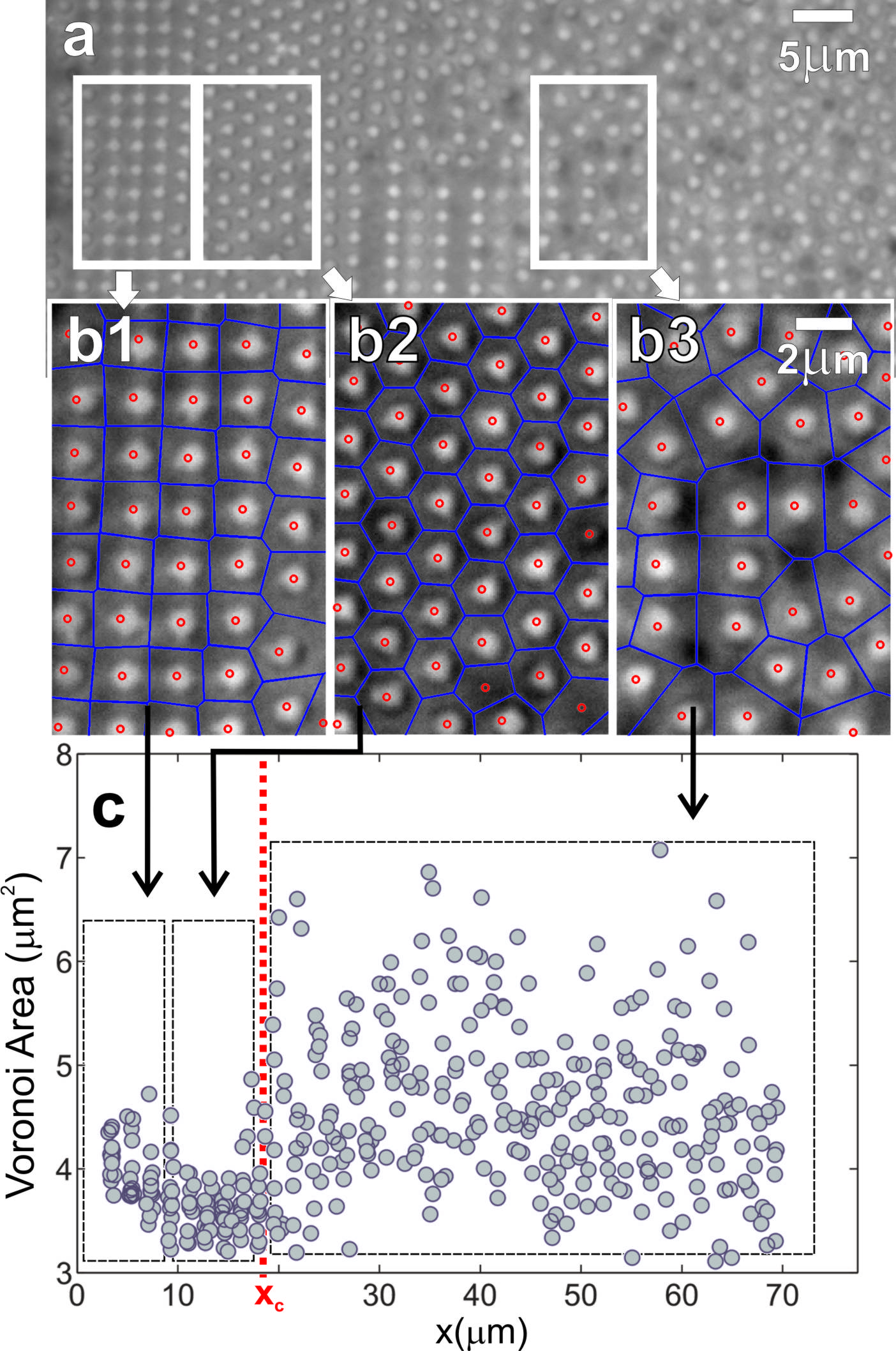}
	
\caption{Analysis of the particle ordering in the stain. (a) Overview of the different patterns observed in the ring stain. The position of the contact line is on the left side of the figure (the first layers of particles are not shown). (b1-3) A close-up reveals the sequence of patterns in the stain: square packing close to the contact line (b1), followed by hexagonal packing (b2), and finally disordered packing (b3). The Voronoi cells belonging to the particles are shown in blue. (c) The area of the Voronoi cells plotted versus the distance $x$ from the contact line. The order-to-disorder transition at $x=x_c$ is determined by the increase in the mean Voronoi area and the dispersion, due to the lack of order. This transition is indicated by the red dotted line.}
\label{fig2}
\end{figure}

When the droplet has completely evaporated, most of the colloidal particles have aggregated into a ring-shaped coffee stain (Fig. \ref{fig1}b). Zooming in on the stain, we see that the particle arrangement is not homogeneous (Fig. \ref{fig1}c): there is a remarkable transition from a crystalline arrangement to a disordered phase. A top view of the stain (Fig. \ref{fig1}d) shows a sequence of particle arrangements within the ordered phase, starting from hexagonal packing, followed by square packing, and again hexagonal packing. The particle arrangement is analyzed in more detail in Fig. \ref{fig2}. Close to the contact line, we find a square arrangement of particles (Fig. \ref{fig2}b1), followed by a hexagonal array (Fig. \ref{fig2}b2), and finally a disordered phase (Fig. \ref{fig2}b3). The same trend has been observed for different particle sizes and concentrations. A well-known method to quantitatively assess the different packing structures, is to calculate the Voronoi areas \cite{Voronoi} of the particles, as shown in Fig. \ref{fig2}b. In Fig. \ref{fig2}c, we identify the Voronoi areas corresponding to the different cell arrangements: the disordered phase exhibits a larger average cell area as well as a wider dispersion, compared to the ordered phase. This reveals the sharp transition from order to disorder.

\begin{figure}
	\includegraphics[width=3.4 in]{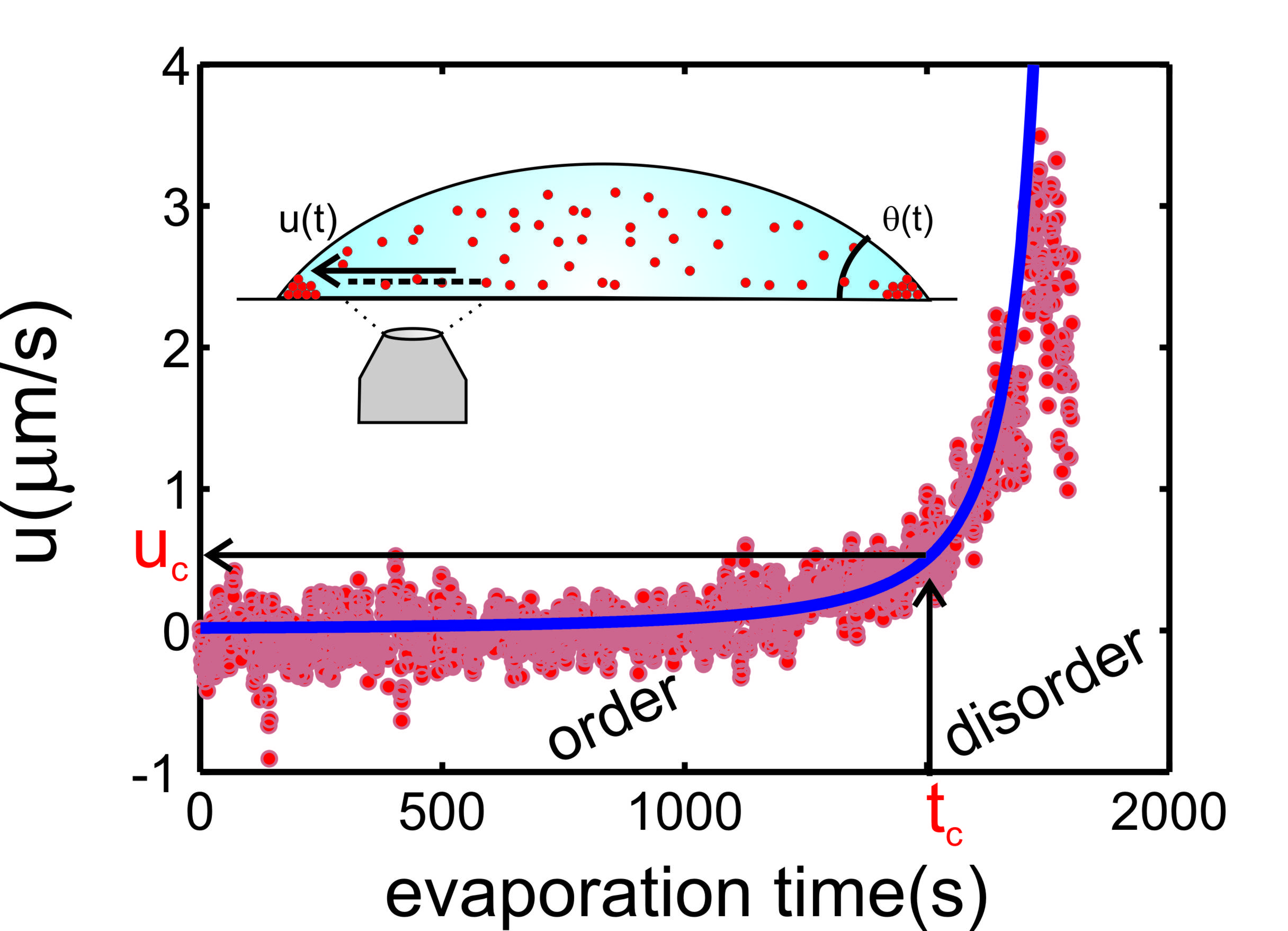}
	
\caption{Origin of the order-to-disorder transition. Plot of the radial particle velocity versus time, measured at a distance of 174 $\mu$m from the contact line. The fluid velocity predicted by the model (\ref{vr}) is plotted as a solid line. Both model and experiment show the dramatic velocity increase at the end of the droplet's life (the rush hour). The inset shows the direction of the radial velocity in the droplet. The time $t_c$ at which the order-to-disorder transition occurs is determined from the videos; the corresponding critical velocity $u_c$ is read from this graph (indicated by the black line in the figure).}
\label{fig3}
\end{figure}

The order-to-disorder transition can be traced back to the hydrodynamics within the drop. Fig. \ref{fig3}a shows the radial velocity measured by $\mu$PIV of the layer of particles closest to the glass slide, in a region close to the contact line. The particle velocity increases dramatically in the last moments of the droplet's life. We refer to this sudden change in speed as ``rush hour''. The particles that arrive early, at a low deposition speed, form an ordered (square or hexagonal) structure. By contrast, particles that arrive during rush hour have a high speed and form a jammed, disordered phase.

The rush-hour behavior can be explained by simple mass-balance considerations. The volume flow towards the contact line inside the drop is driven by the evaporation from the drop surface \cite{Deegan:1997}. It turns out that the rate of evaporation is approximately constant over time \cite{Popov:2005}. To replenish this evaporated liquid, a continuous volume flow towards the contact line is generated inside the drop. However, the drop height is vanishing during evaporation. This can be characterized by a contact angle $\theta(t)\to 0$ at the final stages of evaporation (see Fig. \ref{fig3}). Hence, the same amount of liquid has to be squeezed through an area which is vanishing, inducing a diverging radial velocity.
Close to the contact line, the dependence of the height-averaged radial velocity $\bar{u}$ on time $t$ and distance from the drop center $r$, is given by \cite{Deegan:2000, Popov:2005}
\begin{figure}
	\includegraphics[width=3.4 in]{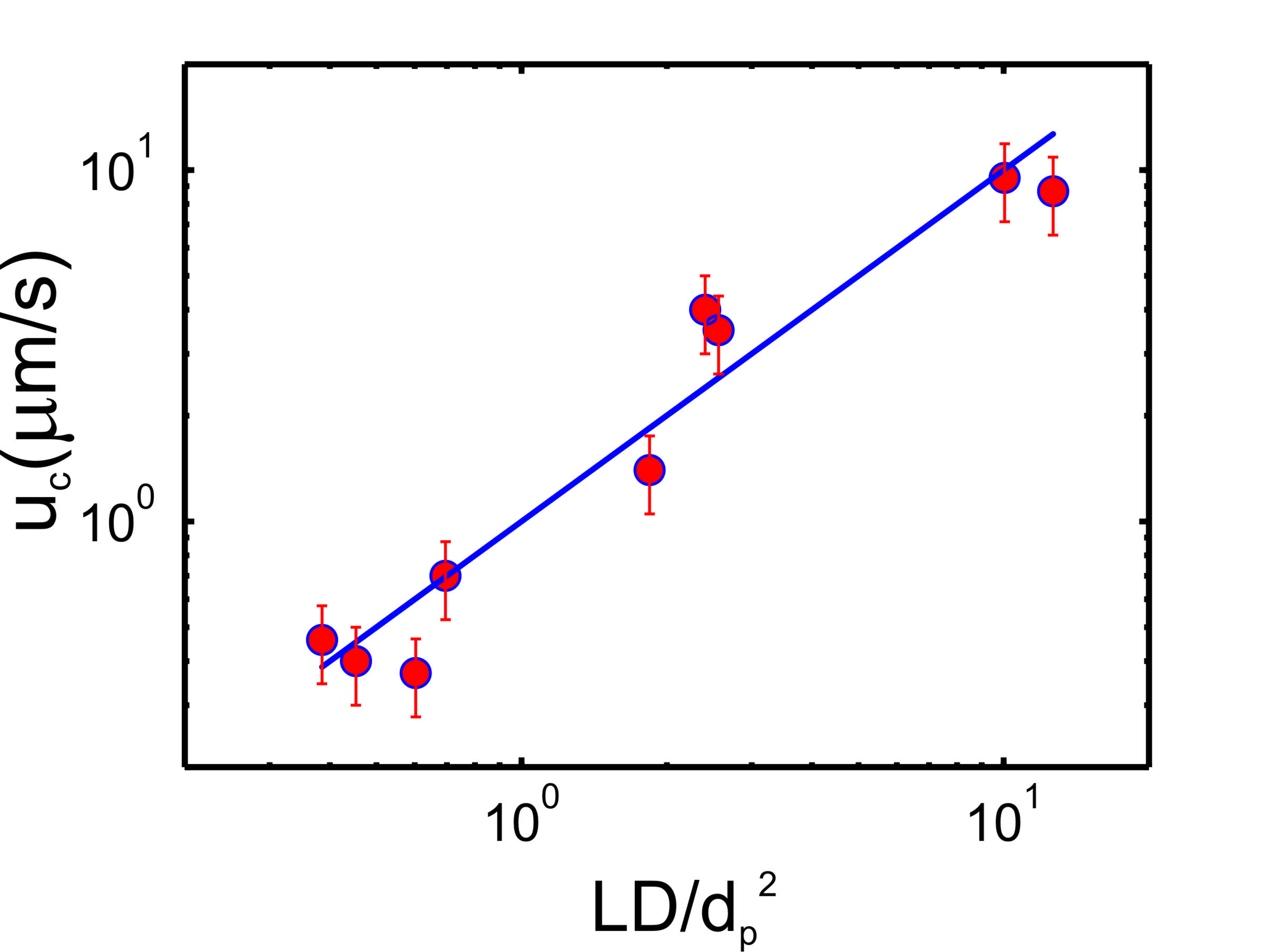}
\caption{The experimentally determined critical velocity $u_c$, versus its theoretical prediction (\ref{uc}).}
\label{fig4}
\end{figure}
\begin{equation}
\bar{u}(r,t)=\frac{D^*}{\theta(t)}\frac{1}{\sqrt{R(R-r)}};\label{vr}
\end{equation}
see \cite{sup} for details.
Here, $D^*=2\sqrt{2}D_{va}\Delta c/(\pi\rho)$ represents the driving of the flow, with $D_{va}=24\times 10^{-6}$ m$^2$/s the diffusion constant for vapor in air, $\Delta c=1.2\times 10^{-2}$ kg/m$^3$ the vapor concentration difference between the drop surface and the surroundings, $\rho =998$ kg/m$^3$ the liquid density, and $R$ the drop radius. The spatial $\left(R-r\right)^{-1/2}$-divergence originates from the diverging evaporative flux from the drop surface \cite{Deegan:1997}. Although this \emph{spatial} divergence generated a lot of interest \cite{Deegan:1997, Deegan:2000, Hu:2005, Popov:2005, bodiguel:2010, mugele}, it is clear from (\ref{vr}) that there is also a \emph{temporal} divergence because $\theta(t)$ vanishes. This temporal divergence provides the key to the rush-hour behavior: the radial velocity blows up towards the end of the droplet's life.

The experimental velocity measurements are performed at a distance of 2 $\mu$m above the glass slide, and therefore differ from the theoretical height-averaged velocity. To quantitatively compare the model with the experimental data, a description of the velocity profile is required. Given the low velocities measured inside the evaporating drop (low Reynolds number) the inertial terms in the Navier-Stokes equation can be neglected. Since the droplet height is much smaller than its radius, the thin-film (lubrication) approximation \cite{Oron:1997} can be used to describe the velocity field. Hence, the velocity $u(r,z,t)$ \cite{sup} is given by
\begin{equation}
u(r,z,t)=\frac{3}{h^2(r,t)}\bar{u}(r,t)\left(h(r,t) z-\frac{1}{2}z^2\right),\label{vprof}
\end{equation}
with $h(r,t)=\theta(t)\left(R^2-r^2\right)/(2R)$, the local droplet height.
As shown in Fig. \ref{fig3}, this model quantitatively describes the rush-hour behavior. For the plot of (\ref{vprof}) we took $R-r=174$ $\mu$m (the distance from the contact line where the $\mu$-PIV measurements are performed), $z=2$ $\mu$m and $t_e$ the droplet's life time determined from the experiment. This means there are no adjustable parameters in the comparison between the model and the experiments.

We show that the structural change in the stain can be entirely attributed to the rush-hour behavior.
The particles that form the ordered phase arrive at the contact line at an early time, when the deposition rate is still low. On the other hand, the particles that belong to the disordered phase arrive during rush hour, at a high deposition rate.
The slow, early arriving particles have time to arrange themselves within the stain by Brownian motion, while those coming in rush hour have no time to find an appropriate spot and are quenched into a disordered phase. We can quantify this transition by comparing the time scale on which particles arrange by diffusion to the time scale of particle deposition. The diffusive time scale is $t_D=d_p^2/D$, where $D$ is the diffusivity of the particles in liquid, calculated by the Einstein relation. The hydrodynamic time scale is $t_h=L/u $ (where $L$  is the typical distance between the particles, dependent on their concentration in the solution $n_p(t)$ as $L=n_p^{-1/3}$, and $u$ is the particle velocity). When $t_D$ and $t_h$ become of the same order, a transition is expected. We therefore define the critical transition velocity as

\begin{equation}
u_c \sim LD/d_p^2. \label{uc}
\end{equation}

We validated this criterion experimentally for various particle concentrations and sizes. First, we determined the time $t_c$ when the ring-shaped stain undergoes the transition (the position where this occurs is $x_c$, as indicated by the red line in Fig. \ref{fig2}c). Next, we used the velocity-time plots (Fig. \ref{fig3}) to find the experimental value of $u(t_c)=u_c$. As demonstrated in Fig. \ref{fig4}, the experimental critical velocity perfectly follows the proposed scaling. The agreement suggests that the prefactor in (\ref{uc}) is actually of order unity.
Our model also explains many results described in the literature \cite{Bigioni_Witten, Pauchard, Limat:1995, Dufresne:2003}, where large-scale crystalline deposits were observed for particle sizes ranging from 6 to almost 50 nm, in hindsight. In this range of particle sizes, the critical velocity required for a disordered phase (\ref{uc}) is unreachable within the droplet, which explains the absence of a disordered phase.

Apart from the order-to-disorder transition, there are also transitions within the ordered phase, from square to hexagonal packing and vice versa (Fig. \ref{fig1}d). In an unconfined system, the most efficient packing lattice is the Hexagonal Close Packing (HCP), as conjectured by Kepler \cite{Kepler1611}.
However, in the evaporating drop the particles are confined in a wedge formed by the glass slide on one side, and the liquid-air interface with contact angle $\theta$ on the other side. It was shown by Pieranski et al. \cite{Pieranski:1983, Pieranski:1984} that such a confinement indeed leads to a sequence of hexagonal and square packing structures, dependent on the most efficient packing for the available space. Indeed, when a new layer is formed, the confinement by the wedge favors the square packing; see Fig 1. Away from the step edge, the available space increases and allows for the denser, hexagonal packing structure. We observe a sequence of this type of transitions, each time a new layer is added. Similar transitions were also observed by Abkarian et al. \cite{HStone2004} in a wedge formed by a moving contact line.

In conclusion, in this Letter we revealed the mechanism triggering crystallization in an evaporating colloidal solution drop with a pinned contact line. The temporal divergence (rush hour) observed in the velocity field inside the drop is responsible for the lack of order in the ring-shaped stains: when the deposition rate is low, particles have time to arrange by Brownian motion and form an ordered phase. During the rush hour, there is no time for such re-arrangement and the particles form a disordered phase. Such effect resembles the popular computer game \emph{Tetris}, where at the beginning of the game slowly falling objects are easily arranged in regular structures, while as the game evolves, the fast falling objects become jammed and disordered.
Therefore, the key to generate large-scale colloidal crystals is to avoid the rush-hour behavior, i.e. to avoid small contact angles during evaporation. This can be achieved for example by modifying the hydrophobicity of the substrate or by continuously replenishing the evaporated liquid.
Large-scale crystalline arrays can also be obtained by tuning the particle size, since nanometer-sized particles diffuse much more efficiently. As a rule of thumb, freely-evaporating colloidal solutions with particles within the nanometric range will naturally crystallize, micrometer-sized particles will show both ordered and disordered phases, while larger particles lead to disordered stains


\begin{acknowledgements}
We gratefully acknowledge Harold J. W. Zandvliet, Peter Schall and Ardi Dortsmans for their suggestions, and D. Arnaldo del Cerro for assistance with the SEM. The financial support of the NWO-Spinoza program is also acknowledged.
\end{acknowledgements}






\begin{thebibliography}{22}%
\makeatletter
\providecommand \@ifxundefined [1]{%
 \@ifx{#1\undefined}
}%
\providecommand \@ifnum [1]{%
 \ifnum #1\expandafter \@firstoftwo
 \else \expandafter \@secondoftwo
 \fi
}%
\providecommand \@ifx [1]{%
 \ifx #1\expandafter \@firstoftwo
 \else \expandafter \@secondoftwo
 \fi
}%
\providecommand \natexlab [1]{#1}%
\providecommand \enquote  [1]{``#1''}%
\providecommand \bibnamefont  [1]{#1}%
\providecommand \bibfnamefont [1]{#1}%
\providecommand \citenamefont [1]{#1}%
\providecommand \href@noop [0]{\@secondoftwo}%
\providecommand \href [0]{\begingroup \@sanitize@url \@href}%
\providecommand \@href[1]{\@@startlink{#1}\@@href}%
\providecommand \@@href[1]{\endgroup#1\@@endlink}%
\providecommand \@sanitize@url [0]{\catcode `\\12\catcode `\$12\catcode
  `\&12\catcode `\#12\catcode `\^12\catcode `\_12\catcode `\%12\relax}%
\providecommand \@@startlink[1]{}%
\providecommand \@@endlink[0]{}%
\providecommand \url  [0]{\begingroup\@sanitize@url \@url }%
\providecommand \@url [1]{\endgroup\@href {#1}{\urlprefix }}%
\providecommand \urlprefix  [0]{URL }%
\providecommand \Eprint [0]{\href }%
\providecommand \doibase [0]{http://dx.doi.org/}%
\providecommand \selectlanguage [0]{\@gobble}%
\providecommand \bibinfo  [0]{\@secondoftwo}%
\providecommand \bibfield  [0]{\@secondoftwo}%
\providecommand \translation [1]{[#1]}%
\providecommand \BibitemOpen [0]{}%
\providecommand \bibitemStop [0]{}%
\providecommand \bibitemNoStop [0]{.\EOS\space}%
\providecommand \EOS [0]{\spacefactor3000\relax}%
\providecommand \BibitemShut  [1]{\csname bibitem#1\endcsname}%
\let\auto@bib@innerbib\@empty
\bibitem [{\citenamefont {Stein}\ and\ \citenamefont
  {Schroden}(2001)}]{colloidalphotonics}%
  \BibitemOpen
  \bibfield  {author} {\bibinfo {author} {\bibfnamefont {A.}~\bibnamefont
  {Stein}}\ and\ \bibinfo {author} {\bibfnamefont {R.}~\bibnamefont
  {Schroden}},\ }\href@noop {} {\bibfield  {journal} {\bibinfo  {journal}
  {Curr. Opin. Solid St. M.}\ }\textbf {\bibinfo {volume} {5}},\ \bibinfo
  {pages} {553} (\bibinfo {year} {2001})}\BibitemShut {NoStop}%
\bibitem [{\citenamefont {Zhang}\ \emph {et~al.}(2008)\citenamefont {Zhang},
  \citenamefont {Maheshwari}, \citenamefont {Chang},\ and\ \citenamefont
  {Zhu}}]{evaporativeDNA}%
  \BibitemOpen
  \bibfield  {author} {\bibinfo {author} {\bibfnamefont {L.}~\bibnamefont
  {Zhang}}, \bibinfo {author} {\bibfnamefont {S.}~\bibnamefont {Maheshwari}},
  \bibinfo {author} {\bibfnamefont {H.}~\bibnamefont {Chang}}, \ and\ \bibinfo
  {author} {\bibfnamefont {Y.}~\bibnamefont {Zhu}},\ }\href@noop {} {\bibfield
  {journal} {\bibinfo  {journal} {Langmuir}\ }\textbf {\bibinfo {volume}
  {24}},\ \bibinfo {pages} {3911} (\bibinfo {year} {2008})}\BibitemShut
  {NoStop}%
\bibitem [{\citenamefont {Deegan}\ \emph {et~al.}(1997)\citenamefont {Deegan},
  \citenamefont {Bakajin}, \citenamefont {Dupont}, \citenamefont {Huber},
  \citenamefont {Nagel},\ and\ \citenamefont {Witten}}]{Deegan:1997}%
  \BibitemOpen
  \bibfield  {author} {\bibinfo {author} {\bibfnamefont {R.~D.}\ \bibnamefont
  {Deegan}}, \bibinfo {author} {\bibfnamefont {O.}~\bibnamefont {Bakajin}},
  \bibinfo {author} {\bibfnamefont {T.}~\bibnamefont {Dupont}}, \bibinfo
  {author} {\bibfnamefont {G.}~\bibnamefont {Huber}}, \bibinfo {author}
  {\bibfnamefont {S.}~\bibnamefont {Nagel}}, \ and\ \bibinfo {author}
  {\bibfnamefont {T.}~\bibnamefont {Witten}},\ }\href
  {http://www.nature.com/nature/journal/v389/n6653/abs/389827a0.html}
  {\bibfield  {journal} {\bibinfo  {journal} {Nature}\ }\textbf {\bibinfo
  {volume} {389}},\ \bibinfo {pages} {827} (\bibinfo {year}
  {1997})}\BibitemShut {NoStop}%
\bibitem [{\citenamefont {Deegan}\ \emph {et~al.}(2000)\citenamefont {Deegan},
  \citenamefont {Bakajin}, \citenamefont {Dupont}, \citenamefont {Huber},
  \citenamefont {Nagel},\ and\ \citenamefont {Witten}}]{Deegan:2000}%
  \BibitemOpen
  \bibfield  {author} {\bibinfo {author} {\bibfnamefont {R.~D.}\ \bibnamefont
  {Deegan}}, \bibinfo {author} {\bibfnamefont {O.}~\bibnamefont {Bakajin}},
  \bibinfo {author} {\bibfnamefont {T.}~\bibnamefont {Dupont}}, \bibinfo
  {author} {\bibfnamefont {G.}~\bibnamefont {Huber}}, \bibinfo {author}
  {\bibfnamefont {S.}~\bibnamefont {Nagel}}, \ and\ \bibinfo {author}
  {\bibfnamefont {T.}~\bibnamefont {Witten}},\ }\href
  {http://link.aps.org/doi/10.1103/PhysRevE.62.756} {\bibfield  {journal}
  {\bibinfo  {journal} {Phys. Rev. {E}}\ }\textbf {\bibinfo {volume} {62}},\
  \bibinfo {pages} {756} (\bibinfo {year} {2000})}\BibitemShut {NoStop}%
\bibitem [{\citenamefont {Hu}\ and\ \citenamefont {Larson}(2005)}]{Hu:2005}%
  \BibitemOpen
  \bibfield  {author} {\bibinfo {author} {\bibfnamefont {H.}~\bibnamefont
  {Hu}}\ and\ \bibinfo {author} {\bibfnamefont {R.~G.}\ \bibnamefont
  {Larson}},\ }\href@noop {} {\bibfield  {journal} {\bibinfo  {journal}
  {Langmuir}\ }\textbf {\bibinfo {volume} {21}},\ \bibinfo {pages} {3963}
  (\bibinfo {year} {2005})}\BibitemShut {NoStop}%
\bibitem [{\citenamefont {Popov}(2005)}]{Popov:2005}%
  \BibitemOpen
  \bibfield  {author} {\bibinfo {author} {\bibfnamefont {Y.~O.}\ \bibnamefont
  {Popov}},\ }\href {http://link.aps.org/doi/10.1103/PhysRevE.71.036313}
  {\bibfield  {journal} {\bibinfo  {journal} {Phys. Rev. {E}}\ }\textbf
  {\bibinfo {volume} {71}},\ \bibinfo {pages} {036313} (\bibinfo {year}
  {2005})}\BibitemShut {NoStop}%
\bibitem [{\citenamefont {Bodiguel}\ and\ \citenamefont
  {Leng}(2010)}]{bodiguel:2010}%
  \BibitemOpen
  \bibfield  {author} {\bibinfo {author} {\bibfnamefont {H.}~\bibnamefont
  {Bodiguel}}\ and\ \bibinfo {author} {\bibfnamefont {J.}~\bibnamefont
  {Leng}},\ }\href@noop {} {\bibfield  {journal} {\bibinfo  {journal} {Soft
  Matter}\ }\textbf {\bibinfo {volume} {6}},\ \bibinfo {pages} {5451} (\bibinfo
  {year} {2010})}\BibitemShut {NoStop}%
\bibitem [{\citenamefont {Eral}\ \emph {et~al.}(2011)\citenamefont {Eral},
  \citenamefont {Augustine}, \citenamefont {Duits},\ and\ \citenamefont
  {Mugele}}]{mugele}%
  \BibitemOpen
  \bibfield  {author} {\bibinfo {author} {\bibfnamefont {H.~B.}\ \bibnamefont
  {Eral}}, \bibinfo {author} {\bibfnamefont {D.~M.}\ \bibnamefont {Augustine}},
  \bibinfo {author} {\bibfnamefont {M.~H.~G.}\ \bibnamefont {Duits}}, \ and\
  \bibinfo {author} {\bibfnamefont {F.}~\bibnamefont {Mugele}},\ }\href@noop {}
  {\bibfield  {journal} {\bibinfo  {journal} {Soft Matter}\ }\textbf {\bibinfo
  {volume} {7}},\ \bibinfo {pages} {4954} (\bibinfo {year} {2011})}\BibitemShut
  {NoStop}%
\bibitem [{\citenamefont {Velikov}\ \emph {et~al.}(2002)\citenamefont
  {Velikov}, \citenamefont {Christova}, \citenamefont {Dullens},\ and\
  \citenamefont {van Blaaderen}}]{Velikov:2002}%
  \BibitemOpen
  \bibfield  {author} {\bibinfo {author} {\bibfnamefont {K.~P.}\ \bibnamefont
  {Velikov}}, \bibinfo {author} {\bibfnamefont {C.~G.}\ \bibnamefont
  {Christova}}, \bibinfo {author} {\bibfnamefont {R.~P.~A.}\ \bibnamefont
  {Dullens}}, \ and\ \bibinfo {author} {\bibfnamefont {A.}~\bibnamefont {van
  Blaaderen}},\ }\href {\doibase 10.1126/science.1067141} {\bibfield  {journal}
  {\bibinfo  {journal} {Science}\ }\textbf {\bibinfo {volume} {296}},\ \bibinfo
  {pages} {106 } (\bibinfo {year} {2002})}\BibitemShut {NoStop}%
\bibitem [{\citenamefont {Harris}\ \emph {et~al.}(2007)\citenamefont {Harris},
  \citenamefont {Hu}, \citenamefont {Conrad},\ and\ \citenamefont
  {Lewis}}]{Harris:2007}%
  \BibitemOpen
  \bibfield  {author} {\bibinfo {author} {\bibfnamefont {D.~J.}\ \bibnamefont
  {Harris}}, \bibinfo {author} {\bibfnamefont {H.}~\bibnamefont {Hu}}, \bibinfo
  {author} {\bibfnamefont {J.}~\bibnamefont {Conrad}}, \ and\ \bibinfo {author}
  {\bibfnamefont {J.}~\bibnamefont {Lewis}},\ }\href {\doibase
  10.1103/PhysRevLett.98.148301} {\bibfield  {journal} {\bibinfo  {journal}
  {Phys. Rev. Lett.}\ }\textbf {\bibinfo {volume} {98}},\ \bibinfo {pages}
  {148301} (\bibinfo {year} {2007})}\BibitemShut {NoStop}%
\bibitem [{\citenamefont {Fan}\ and\ \citenamefont
  {Stebe}(2004)}]{KStebe2004assembly}%
  \BibitemOpen
  \bibfield  {author} {\bibinfo {author} {\bibfnamefont {F.}~\bibnamefont
  {Fan}}\ and\ \bibinfo {author} {\bibfnamefont {K.~J.}\ \bibnamefont
  {Stebe}},\ }\href@noop {} {\bibfield  {journal} {\bibinfo  {journal}
  {Langmuir}\ }\textbf {\bibinfo {volume} {20}},\ \bibinfo {pages} {3062}
  (\bibinfo {year} {2004})}\BibitemShut {NoStop}%
\bibitem [{\citenamefont {Bigioni}\ \emph {et~al.}(2006)\citenamefont
  {Bigioni}, \citenamefont {Lin}, \citenamefont {Nguyen}, \citenamefont
  {Corwin}, \citenamefont {Witten},\ and\ \citenamefont
  {Jaeger}}]{Bigioni_Witten}%
  \BibitemOpen
  \bibfield  {author} {\bibinfo {author} {\bibfnamefont {T.~P.}\ \bibnamefont
  {Bigioni}}, \bibinfo {author} {\bibfnamefont {X.~M.}\ \bibnamefont {Lin}},
  \bibinfo {author} {\bibfnamefont {T.~T.}\ \bibnamefont {Nguyen}}, \bibinfo
  {author} {\bibfnamefont {E.~I.}\ \bibnamefont {Corwin}}, \bibinfo {author}
  {\bibfnamefont {T.~A.}\ \bibnamefont {Witten}}, \ and\ \bibinfo {author}
  {\bibfnamefont {H.~M.}\ \bibnamefont {Jaeger}},\ }\href@noop {} {\bibfield
  {journal} {\bibinfo  {journal} {Nat. Mater.}\ }\textbf {\bibinfo {volume}
  {5}},\ \bibinfo {pages} {265} (\bibinfo {year} {2006})}\BibitemShut {NoStop}%
\bibitem [{\citenamefont {Pauchard}\ \emph {et~al.}(1999)\citenamefont
  {Pauchard}, \citenamefont {Parisse},\ and\ \citenamefont
  {Allain}}]{Pauchard}%
  \BibitemOpen
  \bibfield  {author} {\bibinfo {author} {\bibfnamefont {L.}~\bibnamefont
  {Pauchard}}, \bibinfo {author} {\bibfnamefont {F.}~\bibnamefont {Parisse}}, \
  and\ \bibinfo {author} {\bibfnamefont {C.}~\bibnamefont {Allain}},\ }\href
  {http://link.aps.org/doi/10.1103/PhysRevE.59.3737} {\bibfield  {journal}
  {\bibinfo  {journal} {Phys. Rev. {E}}\ }\textbf {\bibinfo {volume} {59}},\
  \bibinfo {pages} {3737} (\bibinfo {year} {1999})}\BibitemShut {NoStop}%
\bibitem [{\citenamefont {Allain}\ and\ \citenamefont
  {Limat}(1995)}]{Limat:1995}%
  \BibitemOpen
  \bibfield  {author} {\bibinfo {author} {\bibfnamefont {C.}~\bibnamefont
  {Allain}}\ and\ \bibinfo {author} {\bibfnamefont {L.}~\bibnamefont {Limat}},\
  }\href {http://link.aps.org/doi/10.1103/PhysRevLett.74.2981} {\bibfield
  {journal} {\bibinfo  {journal} {Phys. Rev. Lett.}\ }\textbf {\bibinfo
  {volume} {74}},\ \bibinfo {pages} {2981} (\bibinfo {year}
  {1995})}\BibitemShut {NoStop}%
\bibitem [{\citenamefont {Dufresne}\ \emph {et~al.}(2003)\citenamefont
  {Dufresne}, \citenamefont {Corwin}, \citenamefont {Greenblatt}, \citenamefont
  {Ashmore}, \citenamefont {Wang}, \citenamefont {Dinsmore}, \citenamefont
  {Cheng}, \citenamefont {Xie}, \citenamefont {Hutchinson},\ and\ \citenamefont
  {Weitz}}]{Dufresne:2003}%
  \BibitemOpen
  \bibfield  {author} {\bibinfo {author} {\bibfnamefont {E.~R.}\ \bibnamefont
  {Dufresne}}, \bibinfo {author} {\bibfnamefont {E.}~\bibnamefont {Corwin}},
  \bibinfo {author} {\bibfnamefont {N.}~\bibnamefont {Greenblatt}}, \bibinfo
  {author} {\bibfnamefont {J.}~\bibnamefont {Ashmore}}, \bibinfo {author}
  {\bibfnamefont {D.}~\bibnamefont {Wang}}, \bibinfo {author} {\bibfnamefont
  {A.}~\bibnamefont {Dinsmore}}, \bibinfo {author} {\bibfnamefont
  {J.}~\bibnamefont {Cheng}}, \bibinfo {author} {\bibfnamefont
  {X.}~\bibnamefont {Xie}}, \bibinfo {author} {\bibfnamefont {J.}~\bibnamefont
  {Hutchinson}}, \ and\ \bibinfo {author} {\bibfnamefont {D.}~\bibnamefont
  {Weitz}},\ }\href {http://link.aps.org/doi/10.1103/PhysRevLett.91.224501}
  {\bibfield  {journal} {\bibinfo  {journal} {Phys. Rev. Lett.}\ }\textbf
  {\bibinfo {volume} {91}},\ \bibinfo {pages} {224501} (\bibinfo {year}
  {2003})}\BibitemShut {NoStop}%
\bibitem [{sup()}]{sup}%
  \BibitemOpen
  \href@noop {} {\bibinfo  {journal} {See supplemental material at
  http://link.aps.org/supplemental/...}\ }\BibitemShut {NoStop}%
\bibitem [{\citenamefont {Voronoi}(1907)}]{Voronoi}%
  \BibitemOpen
\bibfield  {journal} {  }\bibfield  {author} {\bibinfo {author} {\bibfnamefont
  {G.}~\bibnamefont {Voronoi}},\ }\href@noop {} {\bibfield  {journal} {\bibinfo
   {journal} {Journal f\"ur die Reine und Angewandte Mathematik}\ }\textbf
  {\bibinfo {volume} {133}},\ \bibinfo {pages} {97} (\bibinfo {year}
  {1907})}\BibitemShut {NoStop}%
\bibitem [{\citenamefont {Oron}\ \emph {et~al.}(1997)\citenamefont {Oron},
  \citenamefont {Davis},\ and\ \citenamefont {Bankoff}}]{Oron:1997}%
  \BibitemOpen
  \bibfield  {author} {\bibinfo {author} {\bibfnamefont {A.}~\bibnamefont
  {Oron}}, \bibinfo {author} {\bibfnamefont {S.~H.}\ \bibnamefont {Davis}}, \
  and\ \bibinfo {author} {\bibfnamefont {S.~G.}\ \bibnamefont {Bankoff}},\
  }\href {http://link.aps.org/doi/10.1103/RevModPhys.69.931} {\bibfield
  {journal} {\bibinfo  {journal} {Rev. Mod. Phys.}\ }\textbf {\bibinfo {volume}
  {69}},\ \bibinfo {pages} {931} (\bibinfo {year} {1997})}\BibitemShut
  {NoStop}%
\bibitem [{\citenamefont {Kepler}(1966)}]{Kepler1611}%
  \BibitemOpen
  \bibfield  {author} {\bibinfo {author} {\bibfnamefont {J.}~\bibnamefont
  {Kepler}},\ }\href@noop {} {\emph {\bibinfo {title} {Strena seu de nive
  sexangula, 1611}}}\ (\bibinfo  {publisher} {English translation, Clarendon
  Press, Oxford},\ \bibinfo {year} {1966})\BibitemShut {NoStop}%
\bibitem [{\citenamefont {Pieranski}\ \emph {et~al.}(1983)\citenamefont
  {Pieranski}, \citenamefont {Strzelecki},\ and\ \citenamefont
  {Pansu}}]{Pieranski:1983}%
  \BibitemOpen
  \bibfield  {author} {\bibinfo {author} {\bibfnamefont {P.}~\bibnamefont
  {Pieranski}}, \bibinfo {author} {\bibfnamefont {L.}~\bibnamefont
  {Strzelecki}}, \ and\ \bibinfo {author} {\bibfnamefont {B.}~\bibnamefont
  {Pansu}},\ }\href {http://link.aps.org/doi/10.1103/PhysRevLett.50.900}
  {\bibfield  {journal} {\bibinfo  {journal} {Phys. Rev. Lett.}\ }\textbf
  {\bibinfo {volume} {50}},\ \bibinfo {pages} {900} (\bibinfo {year}
  {1983})}\BibitemShut {NoStop}%
\bibitem [{\citenamefont {Pansu}\ \emph {et~al.}(1984)\citenamefont {Pansu},
  \citenamefont {Pieranski},\ and\ \citenamefont {Pieranski}}]{Pieranski:1984}%
  \BibitemOpen
  \bibfield  {author} {\bibinfo {author} {\bibfnamefont {B.}~\bibnamefont
  {Pansu}}, \bibinfo {author} {\bibfnamefont {P.}~\bibnamefont {Pieranski}}, \
  and\ \bibinfo {author} {\bibfnamefont {P.}~\bibnamefont {Pieranski}},\
  }\href@noop {} {\bibfield  {journal} {\bibinfo  {journal} {J. Physique}\
  }\textbf {\bibinfo {volume} {45}},\ \bibinfo {pages} {331} (\bibinfo {year}
  {1984})}\BibitemShut {NoStop}%
\bibitem [{\citenamefont {Abkarian}\ \emph {et~al.}(2004)\citenamefont
  {Abkarian}, \citenamefont {Nunes},\ and\ \citenamefont {Stone}}]{HStone2004}%
  \BibitemOpen
  \bibfield  {author} {\bibinfo {author} {\bibfnamefont {M.}~\bibnamefont
  {Abkarian}}, \bibinfo {author} {\bibfnamefont {J.}~\bibnamefont {Nunes}}, \
  and\ \bibinfo {author} {\bibfnamefont {H.}~\bibnamefont {Stone}},\
  }\href@noop {} {\bibfield  {journal} {\bibinfo  {journal} {J. Am. Chem.
  Soc.}\ }\textbf {\bibinfo {volume} {126}},\ \bibinfo {pages} {5978} (\bibinfo
  {year} {2004})}\BibitemShut {NoStop}%
\end{thebibliography}
\end{document}